\documentclass[cits]{PoS}
\pdfoutput=0
\title{Effects of Low vs. High Fermionic Modes on Hadron Mass Generation}

\ShortTitle{Effects of Low vs. High Fermionic Modes on Hadron Mass Generation}

\author{\speaker{M. Denissenya}\\
        Institut f\"ur Physik, FB Theoretische Physik, Universit\"at Graz\\
        E-mail: \email{mikhail.denissenya@uni-graz.at}}

\author{L.~Ya.~Glozman\\
        Institut f\"ur Physik, FB Theoretische Physik, Universit\"at Graz\\
        E-mail: \email{leonid.glozman@uni-graz.at}}
        
\author{C.~B.~Lang\\
        Institut f\"ur Physik, FB Theoretische Physik, Universit\"at Graz\\
        E-mail: \email{christian.lang@uni-graz.at}} 
        
 \author{M. Schr\"ock\\
        Institut f\"ur Physik, FB Theoretische Physik, Universit\"at Graz\\
        E-mail: \email{mario.schroeck@uni-graz.at}}

\abstract{A nonvanishing spectral density of the low-lying eigenmodes of the Dirac operator naturally is a signal for dynamical chiral symmetry breaking (D$\chi$SB) via the
Banks-Casher relation. The low-lying eigenmodes alone saturate the pseudoscalar channel and the corresponding propagator successfully reproduces the 
pion mass. In this paper we investigate the effects on the mass generation of hadrons other than pions. The
evolution of these masses upon inclusion of an increasing number of the low-lying eigenmodes is confronted with the hadron mass spectrum upon removal  of such eigenmodes.   
}

\FullConference{31st International Symposium on Lattice Field Theory LATTICE 2013\\
		 July 29 -- August 3, 2013\\
		 Mainz, Germany}

\usepackage[tight]{units}
\usepackage{amsmath}
\usepackage{amssymb}
\usepackage{dsfont}
\usepackage{float}
\renewcommand{\thefigure}{\arabic{figure}}
\renewcommand{\thetable}{\arabic{table}}

\newcommand{\fig}[1]{Fig.~{\ref{#1}}}
\newcommand{\tab}[1]{Tab.~{\ref{#1}}}
\newcommand{\ov}{\overline}
\newcommand{\E}{\mathrm{e}}

\begin{document}

\section{Introduction}

In this contribution we study the question to what extent
the quark condensate of the vacuum (or, better to say, the physics that is responsible
for the dynamical chiral symmetry breaking) contributes also to the
hadron mass generation. Such a question was already addressed on a qualitative
level in lattice studies \cite{DeGrand:2004m,Bali:2010}. One of the conclusions
was that the pion mass is saturated by the physics related to spontaneous
breaking of chiral symmetry, in agreement with general theoretical
expectations. We ask, however, whether it is also so or not for some
other low-lying states, like $\rho$-meson, nucleon and $a_1$-meson.
According to many different model views their mass generation is also related
to the chiral symmetry breaking dynamics. This study is complementary to
our previous work \cite{Lang:2011vw,Glozman:2012fj}, where we investigated what
happens with these hadrons if we artificially restore chiral symmetry
by  removing of the low-lying modes from the quark propagators.

The density of the low-lying Dirac eigenmodes is related to the order parameter of D$\chi$SB, namely,
the quark condensate \cite{Banks:1979yr}. The number of these eigenmodes responsible for the quark condensate scales 
with the lattice volume. We construct hadron propagators from quark propagators considering only a subset of eigenmodes 
of the  spectral representation of  the Dirac operator. The propagator built from $k$ low-lying modes only, we call a truncated 
propagator. A propagator constructed by removing the  low-lying modes from the complete set, we call a reduced 
propagator. The truncated propagator is given by   
 \begin{equation}
  S_{LM(k)}=\sum_{i=1}^{\mathbf{k}}\,\frac{1}{\lambda_i}\,|\lambda_i\rangle \langle \lambda_i|\;, 
 \end{equation}
where $\lambda_i$ and $|\lambda_i\rangle$ are the  eigenvalues and eigenmodes of the chirally improved Dirac
Operator $\gamma_5 D_{CI}$  \cite{Gattringer:2000js}.
By constructing the reduced quark propagator, 
  \begin{equation}
  S_{RD(k)}=S_{Full}-S_{LM(k)}\;, 
 \end{equation}
we get access to the complementary part of the (high-lying) eigenmodes. 
We performed our studies on 160 gauge configurations with $n_f=2$ dynamical fermions with a pion mass 
$M_{\pi}=322(5)$ MeV. The lattice size and the lattice spacing are $16^3\times 32$ and $a=0.144(1)$ fm, respectively. We choose values of $k=32$, $64$, and $128$ where approximately 32
eigenmodes correspond to 30 MeV \cite{Lang:2011vw}.
Furthermore the extended set of interpolators enables us to use the variational method. By constructing the
cross-correlation matrix 
\begin{equation}
C_{ij}(t)=\langle 0|\mathcal{O}_i(t)\mathcal{O}_j^\dagger(0)|0\rangle\;,
\end{equation}
and solving generalized eigenvalue problem    
\begin{equation}
C(t)\vec{\upsilon}_n=\tilde{\lambda}^{(n)}(t)C(t_0)\vec{\upsilon}_n\;,
\end{equation}
we obtain the masses of the ground and possibly excited states from the exponential behaviour of the eigenvalues
$\tilde{\lambda}^{(n)}(t,t_0)= \E^{-E_n (t-t_0)}\left(1+ \mathcal{O}\left(\E^{-\Delta E_n (t-t_0)}\right) \right)$. 
These are extracted for each truncation level $k$ in a given quantum channel. 
The quantum channels and corresponding interpolators are listed in \tab{tab:ints}. 

\begin{table}[t]
  \begin{center}
  \begin{tabular}{|c|c|c|c|c|c|}
    \hline
    \hline

       \#& $\rho$&\# & $a_1$ &\# & $N^{+}$\\ 
    \hline

      8  & $\ov{q}_w \gamma_k\gamma_t q_w$ & 
      1 & $\ov{q}_n \gamma_k \gamma_5 q_n$ &
      1 &$\varepsilon_{abc} q_a (q^T_bC\gamma_5 q_c)$  , $n(nn)$\\ 

      17 & $\ov{q}_{\partial_i} \gamma_k q_{\partial_i}$ & 
      4 & $\ov{q}_w \gamma_k \gamma_5 q_w$ &
      18& $\varepsilon_{abc} i q_a (q^T_bC\gamma_t\gamma_5 q_c)$, $w(ww)$\\
    \hline
  \end{tabular}
  \end{center}
  \caption{Number of interpolators (\#) and the interpolators for the 
             $\rho(1^{--})$,  $a_1(1^{++})$ mesons and
	    the $N^{+}$ nucleon.}\label{tab:ints}
\end{table}
Notations $n$, $w$ and $\partial_i$ stand for the Jacobi smeared -narrow, -wide and derivative sources respectively.
\section{Hadron masses with and without the low modes}\
 We confirmed that the pseudoscalar channel is saturated by the low modes only \cite{DeGrand:2004m}. Here we consider  
 $k=128$ as a sufficient number of the low modes to reproduce the mass of a pion within error bars.  On the other hand when one removes
 the same number of the low modes from the complete set one artificially removes the pion from the spectrum.   
We here determine quantitatively the effects of D$\chi$SB on the mass generation
of hadrons other than the pion expressed in terms of the low-lying eigenmodes. We present our results concerning
$1^{--}$, $1^{++}$, $1/2^+$ channels in the subsequent sections. 
 
\subsection{$\rho(1^{--})$}

\begin{figure}[b]
  \centering
  \begin{tabular}{cc}
    \includegraphics[width=0.45\textwidth]{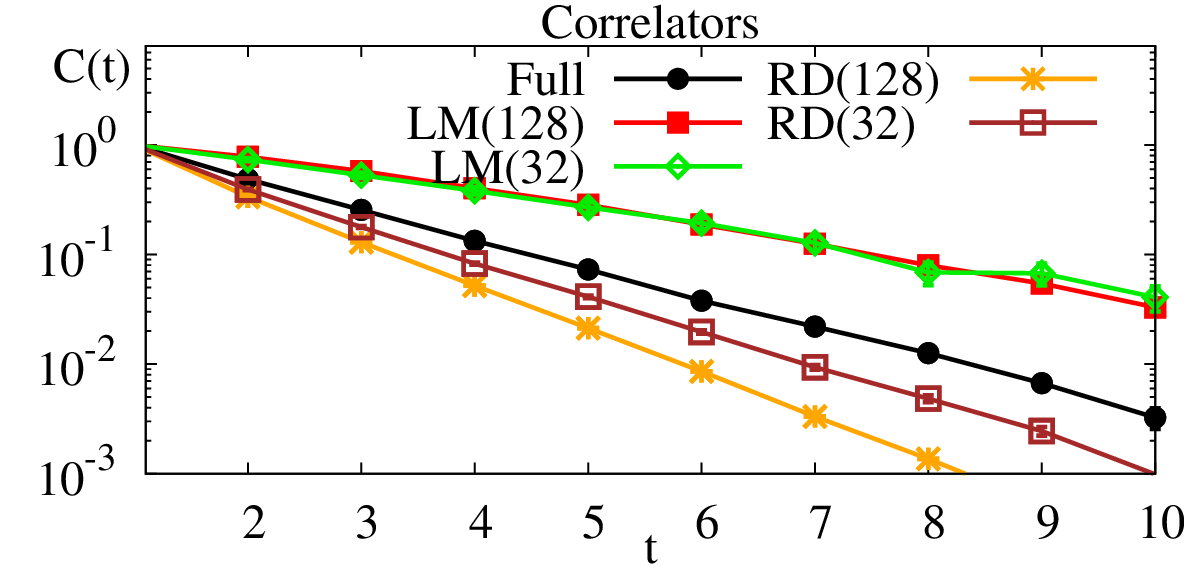} &
    \includegraphics[width=0.45\textwidth]{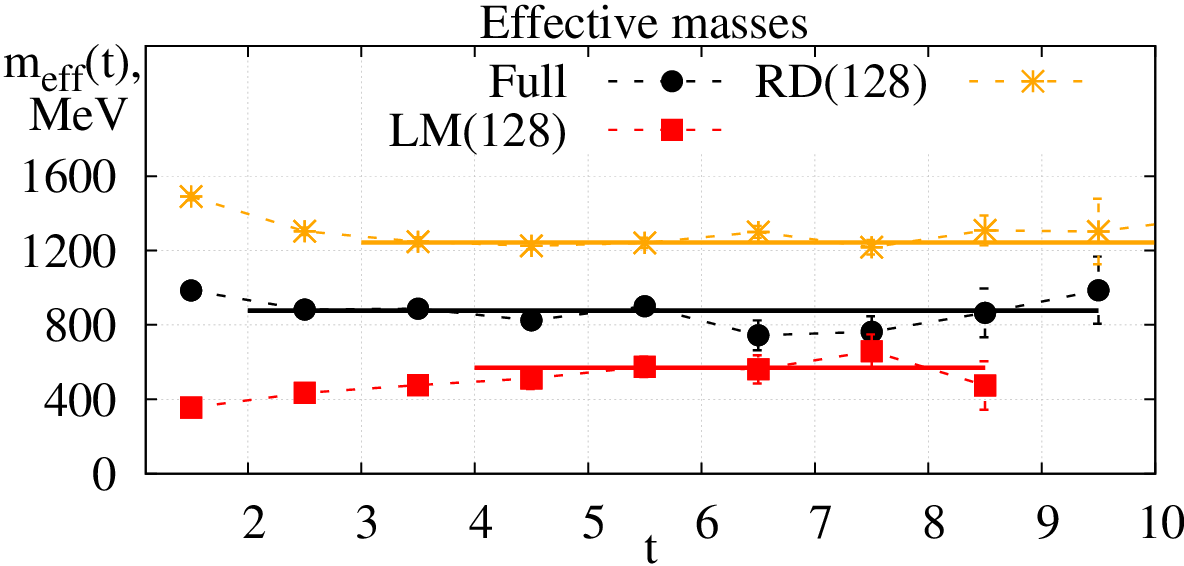}\\
    \includegraphics[width=0.45\textwidth]{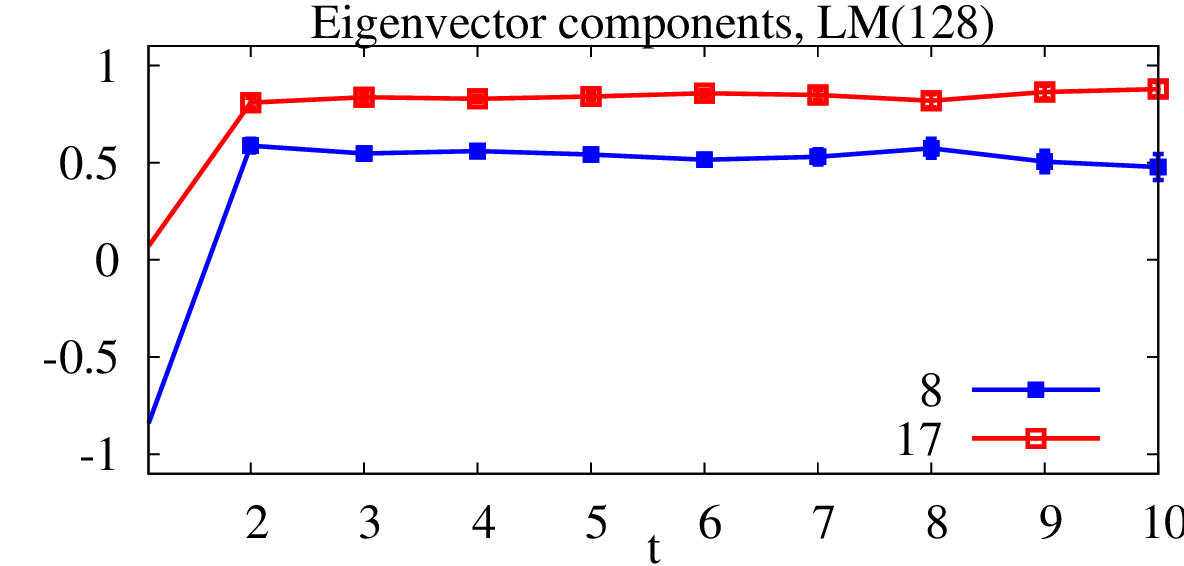} &
    \includegraphics[width=0.45\textwidth]{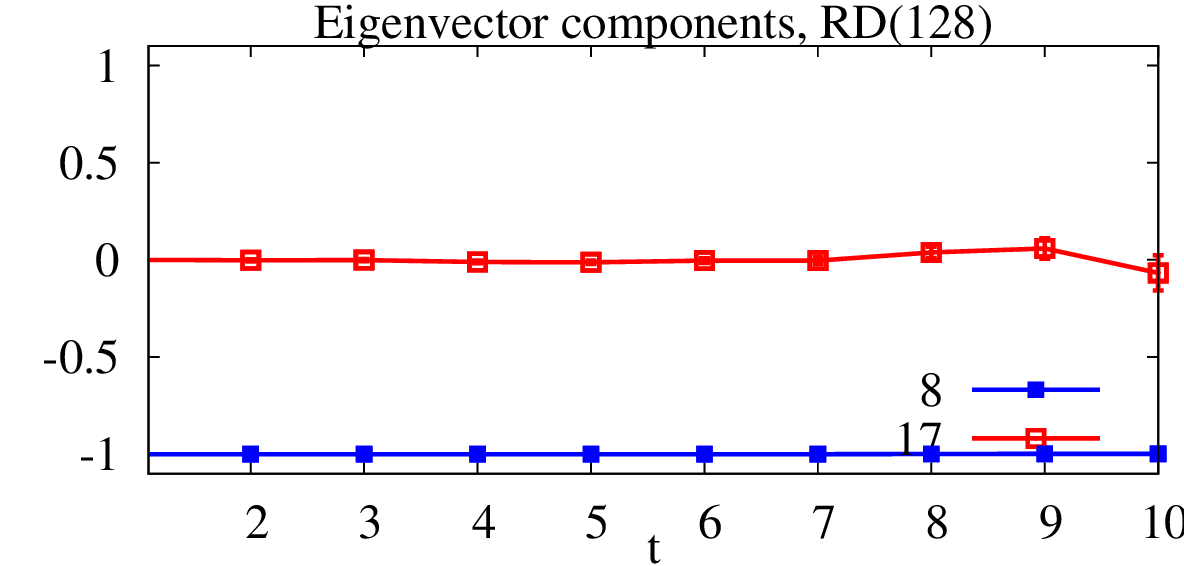}
  \end{tabular} 
   \caption{$\rho$:(top left) correlators, (top right) effective masses,
             (bottom left/right) eigenvectors under the
	     inclusion/removal of the low Dirac eigenmodes} \label{fig:rho}
\end{figure}

We clearly see a bound ground state with only 32 low modes in the $1^{--}$ channel in  \fig{fig:rho}. The decay property 
of the corresponding correlator LM(32) stays almost unchanged upon including 128 low modes into the description.  We 
observe that the mass of this bound state is found to be close to 60 percent of the original full $\rho$ meson mass. On the 
other hand removing 128 low modes results in a larger mass compared to both LM(128) and FULL cases. It is obvious that 
the masses obtained by including the low modes or excluding the low modes are not additive quantities.  

We observe qualitatively the same pattern for the bound state in the $1^{++}$ channel as described above under truncation level $k$. To keep everything consistent we used the same fixed set of interpolators at each truncation level, cf. \fig{fig:rho}. If one
optimises the rather large set of interpolators for each truncation level separately, it is possible to extend the plateau ranges.

\subsection{$N(\frac{1}{2}^{+})$}\label{sec:nuc}

\begin{figure}[t]
  \centering
  \begin{tabular}{cc}
    \includegraphics[width=0.45\textwidth]{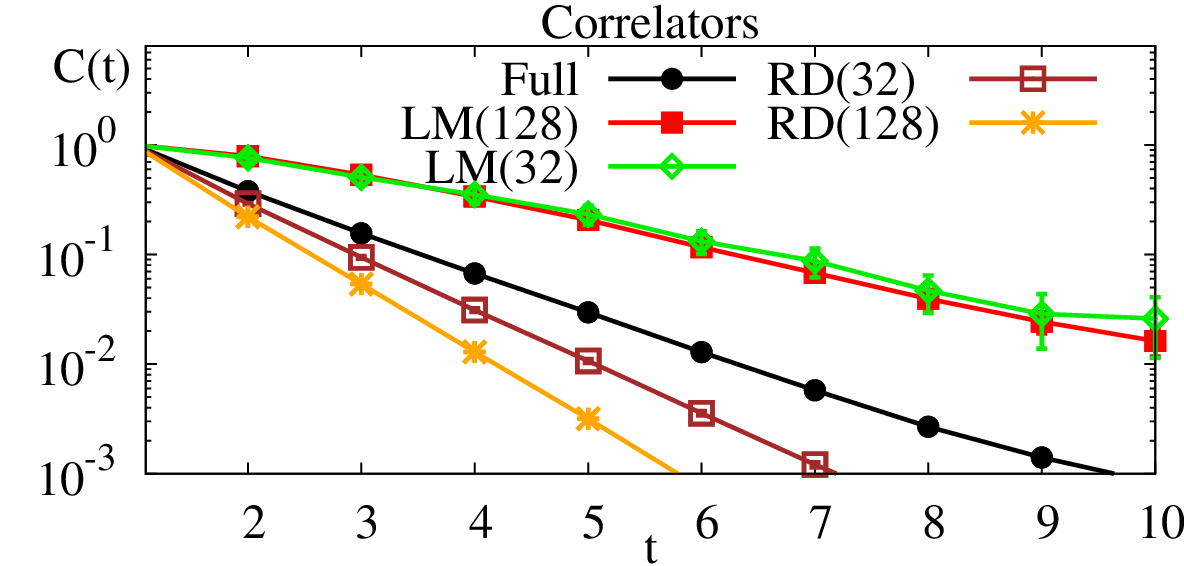} &
    \includegraphics[width=0.45\textwidth]{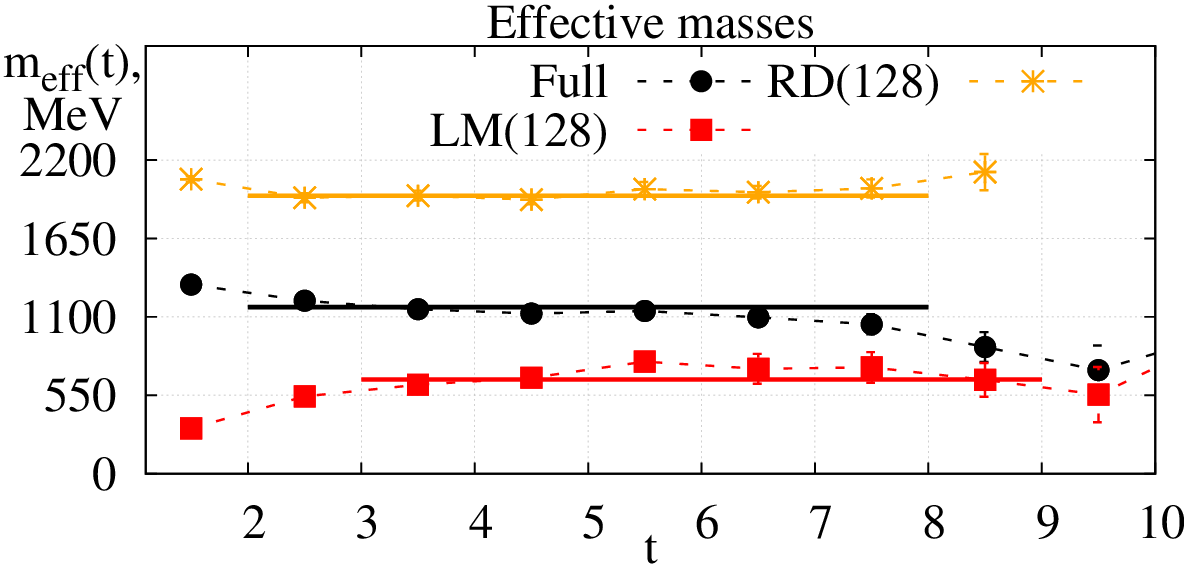}\\
    \includegraphics[width=0.45\textwidth]{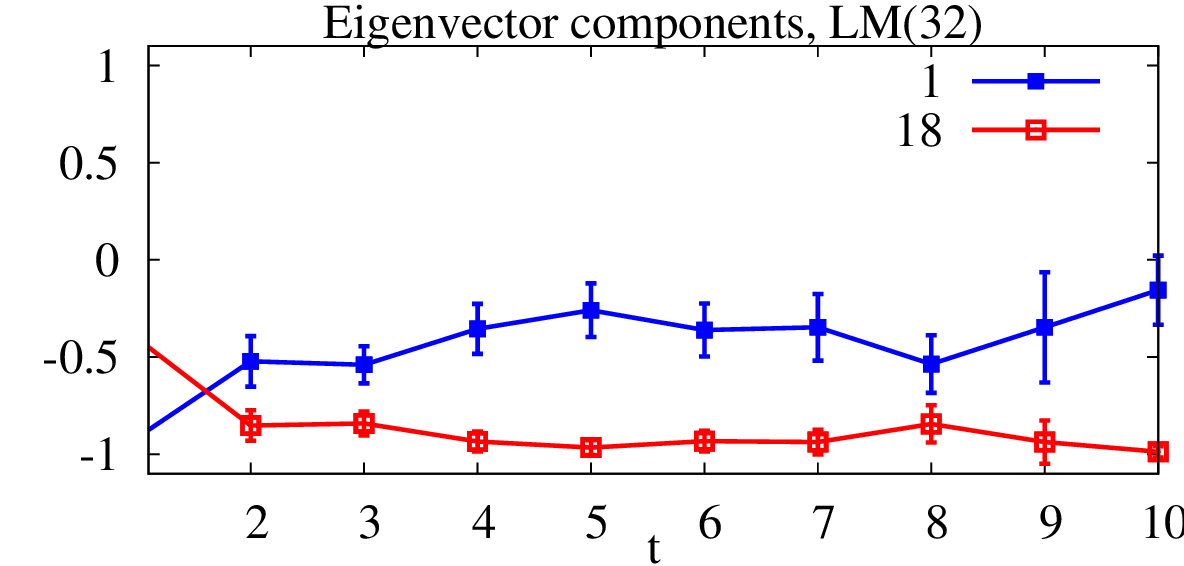} &
    \includegraphics[width=0.45\textwidth]{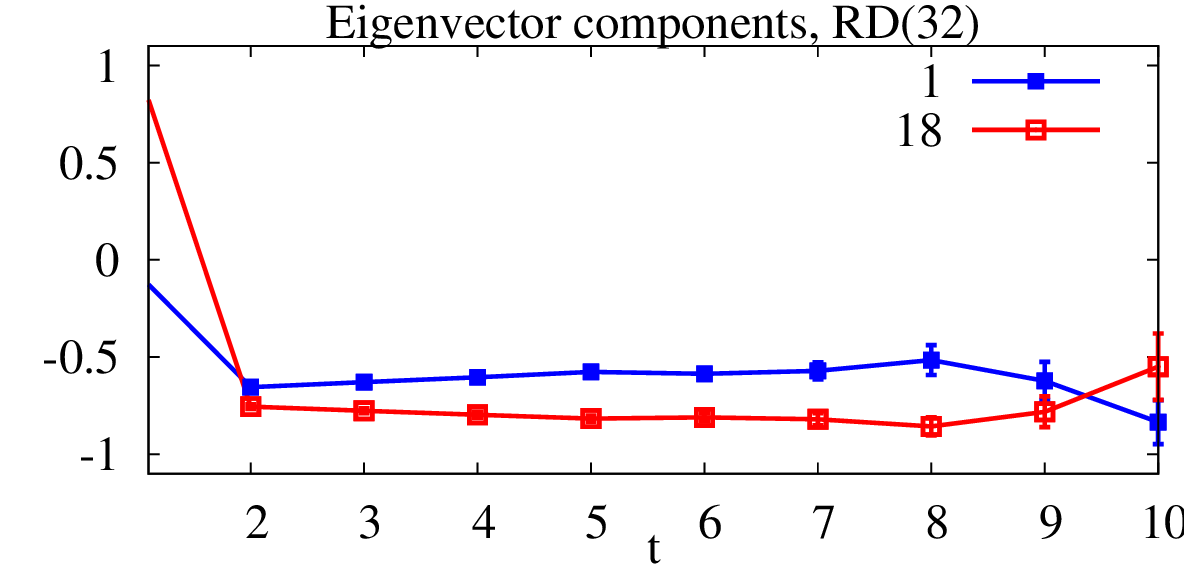}
  \end{tabular} 
  \caption{Nucleon: (top left) correlators, (top right) effective masses,
                    (bottom left/right) eigenvectors under the
		    inclusion/removal of the Dirac eigenmodes}\label{fig:nuc}
\end{figure}

The evolution of the nucleon mass under the low mode inclusion or removal resembles the one we observed for $\rho$ meson bound
state. The contribution of an increasing number of low modes starting from $k=32$  up to $k=128$ does not change
significantly 
as it can be seen from the correlators (\fig{fig:nuc}). Extracting the mass of the nucleon for the case of LM(128) we
recover almost $2/3$ of the full nucleon mass. The low modes provide a large contribution to the nucleon channel of positive parity
but this contribution is not enough to explain the mass of the nucleon without taking into account the effects of the high-lying
eigenmodes. 

\begin{figure}[t]
  \centering
  \begin{tabular}{cc}
    \includegraphics[width=0.45\textwidth]{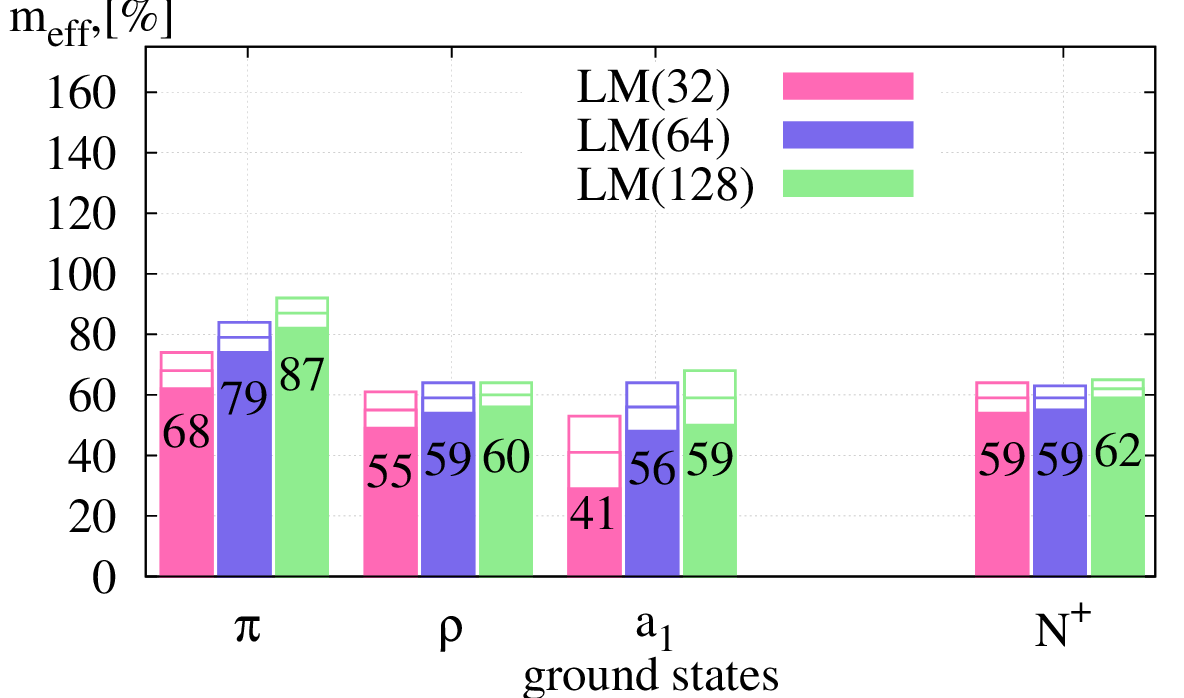}  &
    \includegraphics[width=0.45\textwidth]{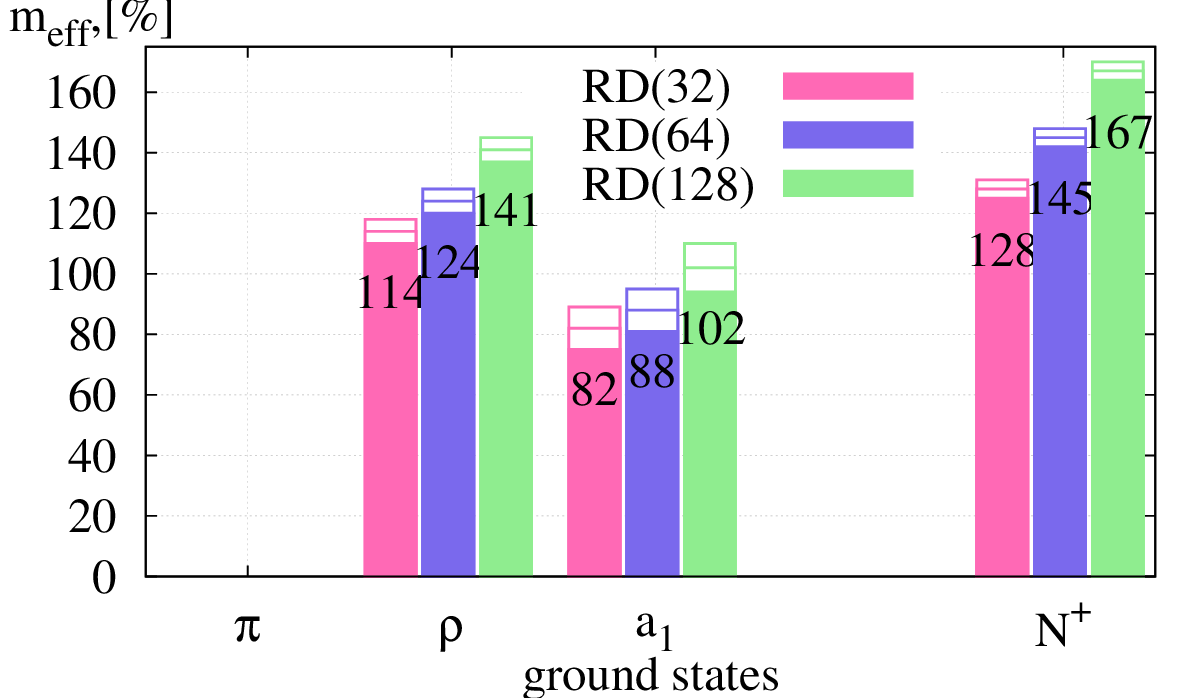} 
 \end{tabular}   
 \caption{Evolution of the hadron masses upon the inclusion (left) and removal (right) of the low eigenmodes} \label{fig:histo}
\end{figure}
\section{Conclusions}

The histograms shown in \fig{fig:histo} illustrate the differences in the hadron mass spectrum depending on whether 
the low-lying modes of the Dirac operator are gradually included or excluded out of the description.   

We conclude that -- unlike the mass of the pion -- the masses of $\rho$, $a_1$, and $N$ grow very slowly with the number of included low modes,
see  \fig{fig:histo} (left). These low modes provide a large contribution to $1^{--}$, $1/2^+$ channels and give a rise up
to $2/3$ of the corresponding full masses. To fill the remaining mass  for the $\rho$-meson and the nucleon one needs to take into account the high-lying eigenmodes.  In contrast, it is the higher-lying modes that are the most essential for the mass of the $a_1$-meson. In particular, it is already sufficient to consider all eigenmodes except for the low-lying ones to obtain the full $a_1$ meson mass, see  \fig{fig:histo} (right). 

\section*{Acknowledgements}
A support from the Austrian Science Fund (FWF) through
the grants DK W1203-N16 and P21970-N16 is acknowledged. 
M.D. is supported by the Austrian Science Fund (FWF) under Grant No. DK W1203-N16. M.S. was supported by the Research Executive Agency (REA) of the European Union under Grant Agreement PITN-GA-2009-238353 (ITN STRONGnet).

\providecommand{\href}[2]{#2}
\begingroup\raggedright

 \endgroup  
\end{document}